\documentclass[final,a4paper,3p]{elsarticle}

\journal{arxiv.org}

\usepackage{array}                   
\usepackage{amsmath,amssymb,amsfonts,amsthm}
\allowdisplaybreaks
\usepackage{dsfont,bm,bbm} 
\usepackage{relsize}
\usepackage{mathrsfs}
\usepackage{booktabs}
\usepackage{mathtools}
\usepackage[german,english]{babel}   
\usepackage{caption}                 
\usepackage[utf8]{inputenc}          
\usepackage[T1]{fontenc}
\usepackage{float}
\usepackage{etoolbox}
\usepackage{graphicx}                
\usepackage{lmodern}                 
\usepackage{microtype}    
\usepackage{lineno}
\usepackage{multirow}
\usepackage{calligra}
\usepackage{nicefrac}
\usepackage[usenames,dvipsnames]{xcolor}
\usepackage{tikz}
\usepackage{tikz-network}
\usepackage[caption=false]{subfig}   
\usepackage{algorithm}
\usepackage[noend]{algpseudocode}

\usepackage[deletedmarkup=xout,authormarkup=none]{changes}
\usepackage{url} 
\usepackage[colorlinks]{hyperref}
\usepackage{cleveref}
\crefformat{equation}{(#2#1#3)}
\AtBeginDocument{%
  \hypersetup{
    citecolor=Black,
    linkcolor=Black,   
    urlcolor=Violet,
	pdfauthor={Uribe et al.}}}


\makeatletter
\def\url@leostyle{%
\@ifundefined{selectfont}{\def\UrlFont{\sf}}{\def\UrlFont{\small\ttfamily}}}
\makeatother
\newcommand*\patchAmsMathEnvironmentForLineno[1]{%
  \expandafter\let\csname old#1\expandafter\endcsname\csname #1\endcsname
  \expandafter\let\csname oldend#1\expandafter\endcsname\csname end#1\endcsname
  \renewenvironment{#1}%
     {\linenomath\csname old#1\endcsname}%
     {\csname oldend#1\endcsname\endlinenomath}}%
\newcommand*\patchBothAmsMathEnvironmentsForLineno[1]{%
  \patchAmsMathEnvironmentForLineno{#1}%
  \patchAmsMathEnvironmentForLineno{#1*}}%
\AtBeginDocument{%
\patchBothAmsMathEnvironmentsForLineno{equation}%
\patchBothAmsMathEnvironmentsForLineno{align}%
\patchBothAmsMathEnvironmentsForLineno{flalign}%
\patchBothAmsMathEnvironmentsForLineno{alignat}%
\patchBothAmsMathEnvironmentsForLineno{gather}%
\patchBothAmsMathEnvironmentsForLineno{multline}%
}

\newcommand{\given}{\;\ifnum\currentgrouptype=16 \middle\fi|\;}
\newcommand{\suchthat}{\;\ifnum\currentgrouptype=16 \middle\fi|\;}

\newcommand{\abs}[1]{\left\vert#1\right\vert}

\newcommand{\ve}[1]{\boldsymbol{#1}}

\graphicspath{{figures/}}

\begin{document}

\begin{frontmatter}

\title{Partially stochastic deep learning with uncertainty quantification for model predictive heating control}

\author[label1,label3]{Emma Hannula\corref{cor1}\fnref{fn1}}
\ead{emma.hannula@lut.fi}
\cortext[cor1]{Corresponding author.}
\fntext[fn1]{These two authors contributed equally to this work.}

\author[label1,label2]{Arttu Häkkinen\fnref{fn1}}

\author[label1,label2]{Felipe Uribe}
\author[label1,label2]{Antti Solonen}
\author[label1,label3]{Jana de Wiljes}
\author[label1]{Lassi Roininen}

\affiliation[label1]{organization={LUT University},
            addressline={Yliopistonkatu 34},
            city={Lappeenranta},
            postcode={53850},
            country={Finland}}

\affiliation[label2]{organization={Danfoss Leanheat},
            addressline={Ilmalantori 1},
            city={Helsinki},
            postcode={00240},
            country={Finland}}

\affiliation[label3]{organization={TU Ilmenau},
            addressline={Ehrenbergstraße 29},
            city={Ilmenau},
            postcode={98693},
            country={Germany}}

\begin{abstract}
Making the control of building heating systems more energy efficient is crucial for reducing global energy consumption and greenhouse gas emissions. Traditional rule-based control methods use a static, outdoor temperature-dependent heating curve to regulate heat input. This open-loop approach fails to account for both the current state of the system (indoor temperature) and free heat gains, such as solar radiation, often resulting in poor thermal comfort and overheating. Model Predictive Control (MPC) addresses these drawbacks by using predictive modeling to optimize heating based on a building's learned thermal behavior, current system state, and weather forecasts. However, current industrial MPC solutions often employ simplified physics-inspired indoor temperature models, sacrificing accuracy for robustness and interpretability. While purely data-driven models offer superior predictive performance and therefore more accurate control, they face challenges such as a lack of transparency.

To bridge this gap, we propose a partially stochastic deep learning (DL) architecture, dubbed LSTM+BNN, for building-specific indoor temperature modeling. Unlike most studies that evaluate model performance through simulations or limited test buildings, our experiments across a comprehensive dataset of 100 real-world buildings, under various weather conditions, demonstrate that LSTM+BNN outperforms an industry-proven reference model, reducing the average prediction error measured as RMSE by more than 40\% for the 48-hour prediction horizon of interest.
Unlike deterministic DL approaches, LSTM+BNN offers a critical advantage by enabling pre-assessment of model competency for control optimization through uncertainty quantification. Thus, the proposed model shows significant potential to improve thermal comfort and energy efficiency achieved with heating MPC solutions.

\end{abstract}


\end{frontmatter}

\section{Introduction and scope}
\label{Intro}
The heating of buildings contributes significantly to global energy consumption and greenhouse gas emissions \cite{IEA_2024, UNEP_2024, Gonzales-Torres_et_al_2022, Urge-Vorsatz_et_al_2015}, especially in northern climates characterized by long and cold winters \cite{Fazeli_et_al_2016}. Global warming and Europe's recent energy crisis further underscore the need for energy-efficient heating solutions.

Water-based central heating systems, often utilizing district heating as the heat source, are prevalent in northern climates, particularly in urban multi-apartment buildings. These systems consist of two networks separated by heat exchangers: the primary side (connected to the heat source) and the secondary side (distributing heat within the building). The secondary side typically includes circuits for domestic hot water and space heating. A heat controller adjusts the primary side flow via control valves to regulate secondary side temperatures. This study focuses on Nordic residential buildings equipped with radiator or underfloor heating systems.

Traditional control methods rely on a deterministic heating curve, adjusting the space heating supply temperature reactively based on measured outdoor temperature. However, this open-loop approach lacks feedback on the system's state (indoor temperature), often leading to inefficiencies through overheating or underheating. It also lacks the ability to anticipate sudden weather changes and account for factors that introduce free heat into the system, such as solar irradiation and internal heat sources (e.g., occupants and electronic devices). The thermal dynamics of buildings are constantly evolving, making the determination of a suitable heating curve a process of endless trial and error, or requiring expensive expert services or complex physics-based white-box modeling.

These shortcomings can be addressed with model predictive control (MPC), which closes the feedback loop by utilizing system state measurements and weather forecasts to plan control actions for the future \cite{Yao_and_Shekhar_2021, Taheri_et_al_2022}. Predictive indoor temperature models formulated to capture thermal dynamics of the building are central to MPC, allowing the inclusion of factors such as solar irradiation and temporal disturbances from internal heat sources. These models adapt to the existing heating dynamics of buildings through automatic updates based on new data. Consequently, MPC provides actual control of the system state, often resulting in reduced energy consumption.

The primary objectives of building heating---thermal comfort and energy efficiency---are generally achieved by maintaining the system state at a predefined target. However, MPC enables the incorporation of more advanced heating optimization objectives. For instance, peak load minimization can be achieved by shifting the heating load based on weather forecasts, predictive modeling, and temporal profiles of domestic hot water usage derived from consumption data. In addition, district heating utilities can leverage large stocks of MPC-controlled buildings as a demand response tool, using them as thermal storage in conjunction with their production optimization. This is feasible because heat loads within buildings can be controlled with greater precision. To realize these benefits, accurate indoor temperature predictions are essential, particularly for longer prediction horizons.

MPC applications can be classified based on the approach to indoor temperature modeling: white-box, gray-box, and black-box models \cite{Drgona_2020}. White-box models are fully physics-based, relying on theoretical principles of heat transfer, energy conservation, and building metadata. Although they offer complete transparency, they are often impractical for scalable solutions. Gray-box models integrate physics-inspired equations with statistical parameter estimation. These models are robust and interpretable, but tend to oversimplify complex dynamics, such as thermal lag effects. Black-box models are purely data-driven alternatives, capable of capturing complex nonlinear relationships when large and descriptive datasets on system inputs and state evolution are available. However, they risk overfitting and lack transparency, necessitating comprehensive datasets and methods to pre-evaluate prediction feasibility to be viable for real-world applications \cite{Hakkinen_2022}.

In industrial heating MPC solutions, the choice of indoor temperature modeling approach can vary based on the available building-specific dataset and evaluations of the models' predictive performance and uncertainty. These evaluations can consider, for instance, how accurately the models capture each building's thermal behavior based on recent data and how uncertain the models are about their predictions given the forecasted weather conditions. Thus, the model with the best potential can be selected for each building and prediction scenario. 

In this work, we aim to develop an accurate probabilistic indoor temperature model that allows for the pre-assessment of model feasibility through predictive uncertainty quantification (UQ). This model can be used in heating MPC for buildings with comprehensive historical datasets across various weather conditions, provided the uncertainty in its predictions is sufficiently low.

\subsection{Related work}
Research in machine learning for time series modeling has primarily focused on exploring Recurrent Neural Networks (RNNs) for specific applications. For instance, Ramadan et al. \cite{Ramadan_2021} applied multiple machine learning methods for indoor temperature prediction and compared their performance against a resistance-capacitance gray-box model. Among these techniques, Long Short-Term Memory (LSTM), a type of RNN, has demonstrated superior accuracy \cite{Xu_2019, Mtibaa_2020, Ma_2022}. Recent studies have aimed to enhance LSTM accuracy by employing methods such as feature extraction with convolutional layers \cite{Elmaz_2021} and incorporating an encoder-decoder architecture with Bayesian hyperparameter optimization \cite{Jiang_2022}. Fang et al. \cite{Fang_2021} highlighted the effectiveness of LSTM models when trained on extensive datasets, such as three years of data. Despite promising results, many of these methods are primarily validated through simulations and lack application to real-world buildings \cite{Bunning_2022}.

A prominent strategy to mitigate the data dependency and generalization issues of purely data-driven building models is to enhance their inductive bias by incorporating prior knowledge of physics. This has led to a significant focus on Physics-Informed Neural Networks (PINNs) \cite{Raissi_2019, Cai_2021} and related architectural hybrids. For comprehensive reviews of this field, we direct the reader to \cite{Hsu_2025, Ma_2025}. Among the most relevant studies, Drgoňa et al. \cite{Drgona_2021} developed a physics-constrained DL model for a multi-zone office building by using the Perron-Frobenius theorem to bound model eigenvalues, thereby improving interpretability and stability. Gokhale et al. \cite{Gokhale_2022} and Pavirani et al. \cite{Pavirani_2024} employed PINNs by embedding physics-based equations directly into the loss function to create control-oriented models. Chen et al. \cite{Chen_2025} studied the sensitivity analysis of physical regularization for a similar approach, where they first estimated the parameters of a resistance-capacitance thermal model from data and then integrated the model into a neural network's constrained loss function. Di Natale et al. \cite{Di_Natale_2022, Di_Natale_2023} combined a linear physics-based module with a neural network in parallel to enforce physics-consistency, though this came at the cost of short-term predictive accuracy. Furthermore, Saeed et al. \cite{Saeed_2024} demonstrated a physics-informed deep reinforcement learning approach for faster, more reliable control decisions to prevent thermal discomfort. While these methods represent substantial contributions, a gap remains for large-scale industrial deployment. Existing approaches often lack: (1) rigorous validation on comprehensive, real-world building datasets; (2) benchmarking against industry-proven reference models (some even fail to surpass the predictive performance of our established baseline); and (3) built-in probabilistic UQ, which is crucial for models used in the control of critical infrastructure like building heating.

Existing applications of UQ in machine learning for time series models encompass several approaches. The Bayesian RNN model discussed in \cite{mirikitani_and_nicolaev_2010}, which employs a sequential approach to update network parameters and hyperparameters, outperforms several traditional time series modeling methods. Another development is the quantile RNN model proposed in \cite{wen_et_al_2017}, which directly learns prediction intervals as lower and upper bounds; this model effectively handles multiple time series, shifting seasonality, future planned event spikes, and cold-starts in large-scale forecasting scenarios. Frequentist approaches to uncertainty estimation are also available. For instance, the method in \cite{alaa_and_vanderschaar_2020} derives predictive uncertainty from the variability of the RNN output's sampling distribution; this is achieved by repeatedly removing sections of temporally correlated training data and collecting predictions from the RNN retrained on the remaining data. Conformal prediction has gained popularity as a frequentist method for constructing distribution-free prediction intervals. However, since this approach requires data exchangeability---an assumption often violated in time series---methods in \cite{xu_and_xie_2021, stankeviciute_et_al_2021} propose techniques to extend its applicability.

\subsection{Contributions and novelty}
While indoor temperature modeling is a well-established field, a significant gap remains in developing models that are simultaneously probabilistic, industrially scalable, and rigorously validated across diverse real-world conditions. This study addresses this gap by making the following key contributions:

\begin{itemize}
    \item[(i)] \textbf{A Probabilistic, Industry-Proven Reference Model for Rigorous Benchmarking.} Existing literature often lacks a robust, industrially-relevant baseline for comparison. We derive and introduce a Bayesian gray-box model that serves as this crucial benchmark. This model is the core of a live MPC solution operating across thousands of buildings in Northern Europe. Its probabilistic nature is essential for handling real-world data noise and providing uncertainty estimates required for model selection. By detailing this model and its computationally efficient posterior inference, we provide a missing foundation for the community, enabling meaningful benchmarking and clarifying the limitations of current industry-standard approaches.
    \item[(ii)] \textbf{A Novel Data-Driven Framework with Built-In UQ for Competency Assessment.} The black-box nature of deep learning (DL) models hinders their adoption in critical systems like building control. While previous research has explored deterministic PINNs, we propose a partially stochastic DL architecture for buildings with comprehensive datasets that provides transparency through UQ. This allows for a crucial feature in pre-assessment of model competency for control optimization---a capability that deterministic models lack. We demonstrate that this UQ is achieved with minimal sacrifice in predictive performance compared to a deterministic variant of our proposed model. Furthermore, we argue that the computational cost of our model is manageable within large-scale MPC architectures, thus addressing a key barrier to the industrial adoption of advanced data-driven models.
    \item[(iii)] \textbf{Comprehensive Real-World Validation Demonstrating Superior Performance.} Moving beyond the limited scope of simulations or single-building tests common in the literature, we conduct a large-scale comparative analysis on a comprehensive dataset from 100 buildings under varied conditions. Our experiments conclusively show that our proposed DL model outperforms the industry-proven reference model across both short- and long-term horizons. This demonstrates that for buildings with sufficient data, implementing our model in an MPC solution---guided by its uncertainty estimates---can directly enhance control, leading to improved comfort and energy efficiency.
\end{itemize}

In summary, this work bridges the gap between academic research and industrial application by focusing on probabilistic, scalable, and rigorously validated modeling. We provide both a robust baseline for future research and a high-performing, uncertainty-aware data-driven alternative, setting a new standard for model development in the context of building heating control.

\subsection{Structure of the paper}
The rest of the paper is organized as follows. Section \ref{Methodology} introduces the reference model and the proposed DL architectures. Section \ref{Experiments} presents the experimental setup, including the dataset and implementation details, and discusses the results of the comparative analysis. Finally, Section \ref{Conclusions} provides concluding remarks and outlines future work.

\section{Methodology}
\label{Methodology}
This section introduces the indoor temperature modeling approaches of our study. As both the reference model and our proposed partially stochastic DL architecture incorporate built-in UQ, we first outline the essential concepts of the Bayesian framework.

\subsection{Bayesian inference}
\label{Bayesian inference}
In the Bayesian framework, model unknowns are represented by probability distributions, inherently incorporating existing uncertainty \cite{Gelman_2013}. Bayesian inference allows for updating beliefs about unknown quantities $\ve{Z}$ in a mathematical model given observed data $\ve{y}\sim \ve{Y}$. Bayesian inference is fundamentally based on Bayes' theorem, with its main components being the \textit{prior}, the \textit{likelihood}, and the \textit{posterior}.
\begin{itemize}
    \item The prior distribution $p(\ve{z})$ expresses initial beliefs, assumptions, or knowledge about the unknown quantities $\ve{Z}$ without observing any data.
    \item The likelihood function $p(\ve{y} \mid \ve{z})$ quantifies the probability of observing the data $\ve{y}$ given a specific realization of the unknown quantity $\ve{Z} = \ve{z}$, reflecting how well the observations support the computational model.
    \item The posterior distribution $p(\ve{z}\mid\ve{y})$ represents the updated beliefs about $\ve{Z}$ after observing $\ve{y}$. It combines the prior assumptions with the new information provided by the data through Bayes' theorem:
    \begin{equation} \label{eq: Bayes' theorem}
    p(\ve{z}\mid\ve{y}) = \frac{p(\ve{y}\mid\ve{z}) p(\ve{z})}{p(\ve{y})},
    \end{equation}
    where $p(\ve{y})=\int p(\ve{y}\mid\ve{z}) \,\mathrm{d} p(\ve{z})$ is the \textit{marginal likelihood} or \textit{evidence}, acting as a normalizing constant.
\end{itemize}

The primary objective of Bayesian inference is to find the posterior probability density $p(\ve{z}\mid \ve{y})$. In practice, this task is often computationally intractable for complex models, primarily due to the high dimensionality of the random variable $\ve{Z}$. As a result, approximations are typically necessary.

\subsubsection{Variational inference}
\label{Variational inference}
Two common probabilistic approaches for approximating the posterior are sampling-based methods, such as Markov Chain Monte Carlo (MCMC) \cite{Gelman_2013}, and an optimization-based method of \textit{variational inference} \cite{Jordan_1998, Bishop_2006, winn_and_bishop_2005, Luttinen_2013}. While MCMC often achieves more precise approximations by drawing a large number of samples from the posterior, variational inference is typically faster, making it better suited for the large-scale industrial application considered in this study.

In variational inference, the core idea is to approximate the true, complex posterior $p(\ve{z}\mid \ve{y})$ as closely as possible with a simpler, tractable distribution $q(\ve{z})$ chosen from a specified family (e.g., a Gaussian) \cite{winn_and_bishop_2005}. This closeness is measured using the Kullback--Leibler (KL) divergence, which quantifies the dissimilarity between two probability distributions:
\begin{equation}
\label{KL}
\begin{split}
    D_{\rm KL}\left(q(\ve{z}) \parallel  p(\ve{z} \mid  \ve{y})\right) &= \int q(\ve{z}) \log \bigg( \frac{q(\ve{z})}{p(\ve{z}\mid  \ve{y})}\bigg) \,\mathrm{d}\ve{z} \geq 0.
\end{split}
\end{equation}

Minimizing the KL divergence directly is challenging because it requires knowledge of the true posterior. Consequently, posterior inference is transformed into a maximization problem of an alternative quantity derived from \eqref{KL}, the evidence lower bound (ELBO) \cite{Jordan_1998, Bishop_2006}:
\begin{equation}
\label{ELBO}
\text{ELBO}(q) = \mathbb{E}_{q} \left[ \log p(\ve{y}|\ve{z}) \right] - D_{\rm KL}\left(q(\ve{z}) \parallel  p(\ve{z})\right).
\end{equation}
This quantity has an intuitive interpretation. The first term, $\mathbb{E}_{q} [ \log p(\ve{y}|\ve{z}) ]$, is the \textit{expected log-likelihood}. Maximizing it encourages a model parameterized by the approximate distribution $q(\ve{z})$ to explain the observed data $\ve{y}$ well. The second term, $-D_{\rm KL}(q(\ve{z}) \parallel p(\ve{z}))$, is the \textit{negative KL divergence} between $q(\ve{z})$ and the prior $p(\ve{z})$. Maximizing it (i.e., minimizing the KL divergence) keeps the approximate posterior $q(\ve{z})$ close to our prior beliefs. Thus, maximizing the ELBO balances fitting the data with maintaining consistency with prior assumptions.

In practice, the maximization problem of ELBO in \eqref{ELBO} with respect to $q(\ve{z})$ can be solved in closed form only for specific variational families, most notably those within the conjugate exponential family. To ensure tractability, a common simplification is the \textit{mean-field approximation} \cite{Bishop_2006}, where the random variables in $\ve{Z}$ are assumed to be independent. Thus, the joint approximate distribution $q(\ve{z})$ factorizes as
\begin{equation}
\label{eq: Mean-field approximation} 
    q(\ve{z}) = \prod_i q(\ve{z}_i).
\end{equation}
While this assumption of independence is strong, it greatly simplifies the computation and is often necessary for practical implementation.

\subsection{Reference model}\label{Reference method}
\begin{figure}[!ht]
    \centering 
    \includegraphics[width=0.75\linewidth]{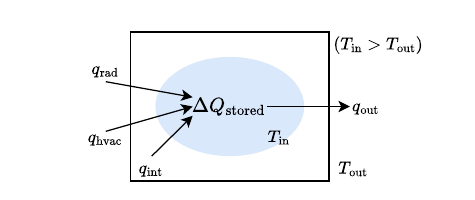}
    \caption{Conservation of thermal energy in a building. Here $\Delta Q_{\text{stored}} = \frac{\text{d}Q_{\text{stored}}}{\text{d}t}$ describes the rate of change of the thermal energy stored by the indoor air mass over time $t$.}
    \label{fig: Energy balance}
\end{figure}

Under steady-state conditions, we consider the air mass inside a building as a system, separated from its surroundings by the building envelope (see Figure \ref{fig: Energy balance}). The system, the envelope, and surroundings are treated separately as homogeneous media with their own uniform thermal properties. Although buildings typically have multiple thermal zones, we simplify the inner volume to a single room. Furthermore, we assume the envelope has minimal thermal capacitance, acting solely as a thermal resistor for heat flow between the system and the environment.

This first-order nodal view of a highly complex and time-varying dynamical system \cite{Ionesi_2019} provides a generalizable basis for scalable MPC solutions without requiring human interference in model calibration. Any building, regardless of its type (e.g. commercial, residential, mixed), topology, or climate zone, can be considered through this simplified view. Although higher-order models would perform well in ideal scenarios, our practical experience indicates that implementing second-order approaches, which consider the envelope's thermal capacitance, presents significant challenges at an industrial scale. These challenges arise primarily due to the introduction of additional latent states and the difficulties in achieving robust parameter estimation from noisy data.

According to the basic physics principle of energy conservation, the change in the thermal energy stored by a system must be equal to the sum of heat flows entering and leaving the system. Applying this and adopting the notation in Figure \ref{fig: Energy balance}, the thermal energy balance equation for a building can be written as:
\begin{equation}
\label{eq: Energy balance equation}
\begin{split}
\frac{\text{d}Q_{{stored}}}{\text{d}t}  = q_{hvac} + q_{out} + q_{rad} + q_{int},
\end{split}
\end{equation}
where $\frac{\text{d}Q_{{stored}}}{\text{d}t}$ is the rate of change of thermal energy stored by the indoor air mass with respect to time, $q_{{hvac}}$ is the heat flow from the heating, ventilation, and air conditioning (HVAC) system, $q_{{out}}$ is the heat flow through the building envelope, $q_{{rad}}$ is the heat flow from solar irradiation, and $q_{{int}}$ is the combined heat flow from various internal heat sources within the building. Here, all right-hand side terms are nonnegative except for $q_{{out}}$, assuming that the indoor temperature is greater than the outdoor temperature.

Let us consider that $q_{{hvac}}$ is brought to the system by a single radiator, which is also a straight pipe, thus assuming the heat flows from other heat sources are negligible. Furthermore, we treat the conductive thermal resistance of the envelope and the convective thermal resistances at its boundary air layers as a single lumped thermal resistor. Expanding the terms in \eqref{eq: Energy balance equation} according to heat transfer equations, rearranging them, and denoting the unknown physical coefficients as the parameter vector $\ve{\theta}\in\mathbb{R}^d$, yields a first-order ordinary differential equation (ODE) for indoor temperature:
\begin{equation}
\label{eq: ODE}
\begin{split}
    \frac{\text{d}T_{{in}}}{\text{d}t} = \theta_1 (T_{{sup}}-T_{{in}}) + \theta_2 (T_{{out}}-T_{{in}}) + \theta_3 \Phi_{{rad}} + \psi(t),
\end{split}
\end{equation}
where $T_{{sup}}$ is the space heating supply water temperature, $T_{{in}}$ is the indoor temperature, $T_{{out}}$ is the outdoor temperature, and $\Phi_{{rad}}$ is the global horizontal irradiation intensity of the sun. For the $q_{{int}}$ term in \eqref{eq: Energy balance equation}, even a simplified physical formula cannot be derived. However, it can be assumed that the effect of this term follows a temporal profile, which can be learned from the data. In this work, the hours of the week are divided into 24 business-day and 24 non-business-day hours (weekends and public holidays), summing up to 48 hourly disturbance profiles for internal sources. Thus, $q_{{int}}$ is written as an indicator function $\psi$ mapping time $t$ to 48 parameters based on the hour of the week. The variables in the other terms are obtained as measured inputs.

Linear state-space models (LSSMs) are extensively used in control theory, time series analysis, and dynamical system modeling \cite{Luttinen_2013, Shumway_2000, Kirk_1998}. Therefore, it is convenient to formulate \eqref{eq: ODE} as an LSSM by including process and observation noise terms, assumed to be Gaussian with zero mean and unknown covariances. This yields the following probabilistic representation for state transition:
\begin{equation}
\label{LSSM}
\begin{split}
    {\widehat{x}}^{(t)} &= (1-\theta_1-\theta_2) {\widehat{x}}^{(t-1)} + \left[\begin{array}{cccc} \theta_1&\theta_2&\theta_3&1  \end{array}\right]\left[\begin{array}{c} T_{{sup}}^{(t)}\\T_{{out}}^{(t)}\\\Phi_{{rad}}^{(t)}\\ \psi(t)  \end{array}\right] + \mathcal{N}(0, \theta_4^{-1}) \\
    {y}^{(t)} &= {\widehat{x}}^{(t)} + \mathcal{N}(0, \theta_5^{-1}),
\end{split}
\end{equation}
where ${\widehat{x}}^{(t)}$ is the estimate of the indoor temperature state at time $t$, and ${y}^{(t)}$ is the indoor temperature measured at time $t$. In this work, the time resolution is hourly, and thus $t-j$ denotes the time at $j$ hours prior to $t$. Collectively, the model, noise, and hourly profile parameters result in the LSSM described in \eqref{LSSM} having a total of $D=53$ parameters. Moreover, let us denote $N$ as the number of total observations, such that $\ve{y}\in\mathbb{R}^N$.

The state and parameter estimation for the LSSM model \eqref{LSSM} is implemented through the Bayesian framework introduced in Section \ref{Bayesian inference}. The joint posterior is expressed as
\begin{equation}
    \label{LSSM posterior}
    p(\ve{\widehat{x}}, \ve{\theta}, \ve{\alpha} \mid \ve{y}) \propto p(\ve{y} \mid \ve{\widehat{x}}, \ve{\theta})
    p(\ve{\widehat{x}} \mid \ve{\theta}) p(\ve{\theta} \mid \ve{\alpha}) p(\ve{\alpha}),
\end{equation}
where $p(\ve{y} \mid \ve{\widehat{x}}, \ve{\theta})$ is the likelihood function after observing the data $\ve{y}$ and $p(\ve{\widehat{x}} \mid \ve{\theta})$ is the state prior probability. Both of these are Gaussian densities derived based on \eqref{LSSM}. Furthermore, we assign conjugate-exponential priors to the remaining terms in \eqref{LSSM posterior} (Gaussian and Gamma distributions, respectively) such that
\begin{equation}
\label{Conjugate priors}
\begin{split}
    p(\ve{\theta} \mid \ve{\alpha}) = \prod_{i=1}^{D} \mathcal{N}(\theta_i \mid  0, \alpha_i^{-1}) \quad\quad p(\ve{\alpha}) = \prod_{i=1}^{D} \mathcal{G}(\alpha_i\mid a, b),
\end{split}
\end{equation}
where the hyperparameters $a$ and $b$ are endowed with non-informative (Uniform) priors.

Due to the model's application in large-scale industrial settings, employing sampling-based methods like MCMC for posterior inference is impractical. This is primarily due to the complex geometry of the posterior and the fact that evaluating the posterior for a single sample requires a complete solution of the Kalman filter \cite{Luttinen_2013, Särkkä_2023}. Consequently, we seek an approximate solution of the posterior in \eqref{LSSM posterior} using variational inference as described in Section \ref{Variational inference}, by assuming the following factorization with respect to the model unknowns:
\begin{equation}\label{Conjugate variational approximation}
p(\ve{z} \mid \ve{y}) \approx q(\ve{z}) = q(\ve{\widehat{x}}) q(\ve{\theta}) q(\ve{\alpha}),
\end{equation}
where we denote $\ve{z}=\{ \ve{\widehat{x}}, \ve{\theta}, \ve{\alpha} \}$ as a set of all unknowns in \eqref{LSSM posterior}. Each component of the factorized variational approximation in \eqref{Conjugate variational approximation} has the same family as the prior on the respective unknown variable, with its own reference parameters. To find these parameters, the ELBO in \eqref{ELBO} is maximized with respect to the posterior approximation in \eqref{Conjugate variational approximation} via the Variational Message Passing (VMP) algorithm \cite{winn_and_bishop_2005}. Given our conjugate-exponential assumptions, VMP offers a computationally efficient solution for the posterior approximation, while still offering a full probabilistic framework for state and parameter estimation, unlike methods such as the maximum a posteriori estimator. Both the VMP method for parameter estimation and the Kalman filter updates for the reference LSSM model are implemented using the software package \texttt{BayesPy}, introduced in \cite{Luttinen_2016}.

\subsection{Deep learning models}
\label{Black-box models}
Deep learning (DL), a subset of machine learning, employs artificial neural networks comprising multiple layers to model intricate patterns in data.  
DL architectures learn hierarchical feature representations by transforming input data through multiple computational layers \cite{LeCun_2015}. 
Model parameters are optimized through iterative training procedures, typically using large-scale datasets and gradient-based optimization methods.

This work focuses on supervised learning, in which models are trained to make predictions based on labeled input-output pairs: $\{\ve{X}, \Delta\ve{y}\}$. Our inputs are defined as $\ve{X}\in \mathbb{R}^{N \times L \times M}$, where $N$ is the number of samples, $L$ is the sequence length (the number of preceding historical instances considered for the input variables given time $t$), and $M$ is the number of input variables. Similarly to our reference model, the input variables of the DL models implemented in this work (presented in Table \ref{tab: Input}) are designed to incorporate physics-based feature engineering. Therefore, instead of dedicating one input dimension to each raw variable, some of the variables are combined to reflect their physical correlations related to the thermodynamic system: $T_{{sup}} - T_{{in}}$ and $\Phi_{{rad}}$ quantify the heat brought to the building by the heating system and solar irradiation, while $T_{{out}} - T_{{in}}$ describes the heat that leaks through the building's envelope into the environment. The disturbance effect from internal heat sources is modeled by using a function $\xi$, which simply maps time $t$ to an integer value based on the hour of the week: 1--24 for non-business-day and 25--48 for business-day hours. Additionally, the azimuth and elevation angles of the sun, $\alpha_{azi}$ and $\alpha_{ele}$, are included, calculated based on the building's GPS coordinates using the software package \texttt{pvlib} \cite{pvlib2018, pvlib2023}. These two additional variables enhance the modeling of the solar effect based on seasonality and building orientation. Finally, to leverage the ability of DL models to capture complex non-linear relationships such as thermal lag effects, each input sequence matrix $\ve{X}^{(t)} = [\ve{x}^{(t-6)}, \ldots, \ve{x}^{(t)}]\in \mathbb{R}^{L \times M}$ in $\ve{X}$ contains $L=7$ most recent hourly observations of the $M=6$ input variables listed in Table \ref{tab: Input}. The length of the input sequence $L=7$ was selected based on experiments with various values ranging from $1$ to $48$.
\begin{table}[!ht]
    \caption{Input variables of the DL models.}
    \centering
    \begin{tabular}{c  c  c}
    \hline
        Variable & Description & Unit \\
        \hline
        $T_{{sup}} - T_{{in}}$ & $\Delta T$ between supply water and inside air & $^\circ$ C\\

        $T_{{out}} - T_{{in}}$ & $\Delta T$ between outside and inside air & $^\circ$ C \\

        $\Phi_{{rad}}$ & Global horizontal irradiation of the sun & W $/$ m$^2$ \\

        $\alpha_{{ele}}$ & Elevation angle of the sun & $^\circ$\\

        $\alpha_{{azi}}$ & Azimuth angle of the sun & $^\circ$ \\

        $\xi(t)$ & An integer value based on the hour of the week & - \\
        \hline
    \end{tabular}

    \label{tab: Input}
\end{table}

To incorporate the physics of the modeled system given the input variables and their connection to \eqref{eq: ODE}, the DL models are trained to predict the change in indoor temperature $\Delta y^{(t)} = T_{{in}}^{(t+1)} - T_{{in}}^{(t)}$ instead of indoor temperature directly. Our previous experience indicates that for machine learning models, this approach yields more precise and physically meaningful predictions \cite{Hakkinen_2022}. Thus, the outputs $\Delta\ve{y}\in\mathbb{R}^N$ represent the differences between the indoor temperature measurements of successive hours. As mentioned in Section \ref{Reference method}, the time resolution of the observed data used in this work is hourly.

\subsubsection{Deterministic deep learning model}\label{sec: LSTM}
Among neural network architectures, RNNs are the most commonly used for modeling sequential data. They differ from standard feedforward networks in their ability to incorporate feedback loops, allowing them to utilize previous model states in their computations \cite{Goodfellow_et_al_2016}. However, standard RNNs often face challenges such as vanishing or exploding gradients \cite{Hewamalage_2021R}. To address these issues, LSTMs were developed, incorporating internal gating mechanisms that allow the learning of long-term dependencies in sequential data \cite{Hochreiter_and_Schmidhuber_1997}. LSTMs are specifically designed to mitigate the vanishing gradient problem by employing a technique called adaptive forgetting to reset model memory at appropriate time intervals \cite{Gers_et_al_2000}.

The structure of an LSTM layer consists of $L$ hidden LSTM units that are connected recursively. Each unit has a forget gate $\ve{f}^{(t)}$, an input gate $\ve{i}^{(t)}$, and a candidate memory gate $\ve{q}^{(t)}$, collectively responsible for managing the model memory, the cell state $\ve{c}^{(t)}$, by determining which stored information should be discarded and which new information should be inserted. The output gate $\ve{o}^{(t)}$ is needed when combining the information contained in the output (hidden state representation) of the previous LSTM unit $\ve{h}^{(t-1)}$, the current input to the unit $\ve{x}^{(t)}$, and the updated cell state $\ve{c}^{(t)}$. The mathematical formulas for the gate operations are given as:
\begin{equation}
    \label{eq: LSTM1}
    \begin{split}
        \ve{f}^{(t)} &= \sigma (\ve{W}_{xf} \ve{x}^{(t)} + \ve{W}_{hf} \ve{h}^{(t-1)} + \ve{b}_f) \\
        \ve{i}^{(t)} &= \sigma (\ve{W}_{xi} \ve{x}^{(t)} + \ve{W}_{hi}\ve{h}^{(t-1)} + \ve{b}_i) \\
        \ve{q}^{(t)} &= \tanh (\ve{W}_{xq}\ve{x}^{(t)} + \ve{W}_{hq}\ve{h}^{(t-1)} + \ve{b}_q) \\
        \ve{o}^{(t)} &= \sigma (\ve{W}_{xo} \ve{x}^{(t)} + \ve{W}_{ho} \ve{h}^{(t-1)} + \ve{b}_o), \\
    \end{split}
\end{equation}
where $\sigma$ denotes the sigmoid activation function, $\ve{W}_{xf}, \ve{W}_{xi}, \ve{W}_{xq}, \ve{W}_{xo}, \ve{W}_{hf}, \ve{W}_{hi}, \ve{W}_{hq}, \ve{W}_{ho}$ are weight parameters, and $\ve{b}_f, \ve{b}_i, \ve{b}_o, \ve{b}_q$ are bias parameters. Cell state $\ve{c}^{(t)}$ and hidden state $\ve{h}^{(t)}$ are then updated by:
\begin{equation}
    \label{eq: LSTM2}
    \begin{split}
        \ve{c}^{(t)} &= \ve{f}^{(t)}\ve{c}^{(t-1)} + \ve{i}^{(t)}\ve{q}^{(t)} \\
        \ve{h}^{(t)} &= \ve{o}^{(t)}\tanh(\ve{c}^{(t)}).        
    \end{split}
\end{equation}

\begin{figure}[!ht]
    \centering
    \includegraphics[width=\linewidth]{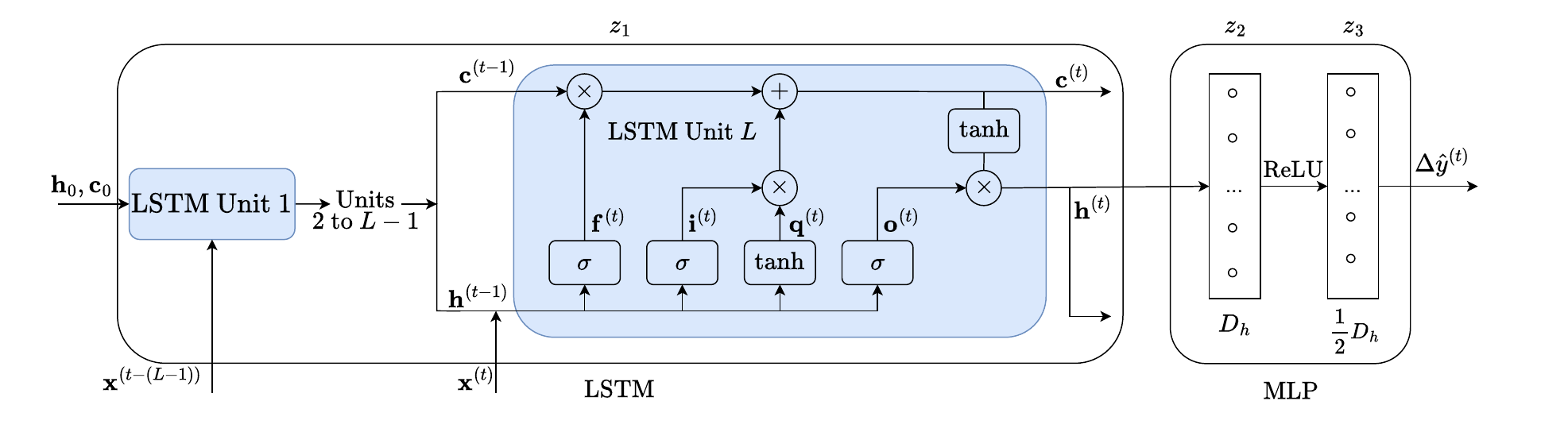}
    \caption{Illustration of an LSTM layer with $L$ LSTM units connected to an MLP with two linear layers. The weights and biases in the LSTM are represented in stacked form as the vector $\ve{z}_1$. Similarly, the weights and biases of the MLP layers are grouped into $\ve{z}_2$ and $\ve{z}_3$, respectively.}
    \label{fig: LSTM}
\end{figure}

We construct our deterministic DL model architecture as follows. An LSTM layer encodes the information in an input sequence $\ve{X}^{(t)} = [\ve{x}^{(t-(L-1))}, \dots, \ve{x}^{(t)}] \in \mathbb{R}^{L \times M}$ into hidden states $[\ve{h}^{(t-(L-1))}, \dots, \ve{h}^{(t)}] \in \mathbb{R}^{L \times D_h}$. The combined information in the final hidden state representation $\ve{h}^{(t)}$ is decoded into a scalar prediction $\Delta \hat{y}^{(t)}$ by a multilayer perceptron (MLP) \cite{Goodfellow_et_al_2016}. The MLP block is built from two linear layers connected by the ReLU activation function. The first linear layer has an input dimension of $D_h$, and the second (output) layer has an input dimension of $\frac{1}{2}D_h$. Based on previous experiments with different hidden dimensionalities, $D_h = 2^{10}$ was selected \cite{Hakkinen_2022}. This model architecture, which we refer to as \emph{LSTM+MLP}, is visualized in Figure \ref{fig: LSTM}.

Mathematically, we can write the forward action of the model given the inputs as
\begin{equation}\label{eq: LSTM+MLP}
\Psi(\ve{X};\ve{z}) = \left(\Psi^{\rm LSTM}(\ve{z}_1) \circ \Psi^{\rm MLP}(\ve{z}_2, \ve{z}_3) \right)(\ve{X}),
\end{equation}
where $\Psi^{\rm LSTM}(\ve{z}_1)$ denotes the LSTM, and $\Psi^{\rm MLP}(\ve{z}_2, \ve{z}_3)$ denotes the MLP computation. The parameters of the LSTM layer are denoted as $\ve{z}_1$, the parameters for the first linear layer are denoted as $\ve{z}_2$, and the parameters for the second linear layer used as the output layer are denoted as $\ve{z}_3$. Thus, all model parameters can be collectively expressed as $\ve{z}=[\ve{z}_1,\ve{z}_2,\ve{z}_3]$.

Finally, the loss function used in LSTM+MLP \eqref{eq: LSTM+MLP} training is the mean absolute error (MAE):
\begin{equation}\label{eq: MAE}
\text{MAE} = \frac{1}{N} \sum_{t} \abs{\Delta y^{(t)} - \Delta \hat{y}^{(t)}},
\end{equation}
where $\Delta \hat{y}^{(t)} =\Psi(\ve{X}^{(t)};\ve{z})$ is the prediction scalar computed by the model at a given time instance $t$.

\subsubsection{Partially stochastic deep learning model}\label{LSTM+BNN}
\textit{Bayesian Neural Networks} (BNNs) extend conventional neural networks by treating the model parameters (weights and biases) as random variables rather than fixed values. By placing prior distributions on these parameters and using Bayes' theorem \eqref{eq: Bayes' theorem} to compute posterior distributions given data, BNNs naturally provide UQ for both their parameters and predictions \cite{Neal_2012}.

Making all parameters in a DL model stochastic significantly increases the computational cost due to the higher dimensionality of the optimized parameter space. However, research by Sharma et al. \cite{Sharma_2023} shows that expressive predictive distributions can be achieved through partially stochastic networks, where only a subset of parameters are treated probabilistically. This approach maintains the benefits of UQ while being more computationally efficient. 

Nevertheless, careful selection of the stochastic parameters is crucial, as not all architectures may meet the criteria for effective probabilistic prediction. There are several alternatives on how to do this selection, and a common method involves isolating a specific layer and treating all its parameters as stochastic.

Our partially stochastic architecture, which we call \textit{LSTM+BNN}, builds upon the deterministic LSTM+MLP model (see Section \ref{sec: LSTM}). Through experimentation with different layer configurations, we found that making only the first linear layer stochastic provided the best balance of predictive performance, expressivity, and computational efficiency. Thus, in our LSTM+BNN architecture, the first linear layer parameters ($\ve{z}_2$) are treated as stochastic, denoted as $\ve{Z}_R$. The LSTM and the second linear layer parameters remain deterministic, collectively denoted as $\ve{z}_D = [\ve{z}_1, \ve{z}_3]$. The resulting partially stochastic model has the form:
\begin{equation}\label{eq: LSTM+BNN}
\Psi(\ve{X}; [\ve{Z}_R,\ve{z}_D]) = \left(\Psi^{\rm LSTM}(\ve{z}_1) \circ \Psi^{\rm MLP}([\ve{Z}_R,\ve{z}_3])\right)(\ve{X}).
\end{equation}

The training of the model \eqref{eq: LSTM+BNN} requires the estimation of the posterior distribution for the random parameters $\ve{Z}_R$. However, computing the exact posterior in this case is intractable. Following the variational inference approach (described in Section \ref{Variational inference}), we approximate the true posterior with a tractable density $q(\ve{z}_R) = \mathcal{N}(\ve{\mu}_R, \ve{\sigma}_R^2\ve{I})$, parameterized by mean $\ve{\mu}_R$ and variance $\ve{\sigma}_R^2$. We maximize the ELBO, which can be written for our model as:
\begin{equation}\label{eq: LSTM+BNN ELBO}
\text{ELBO}(q) = \mathbb{E}_q \bigg[ \left\Vert \Delta \ve{y}-\Psi(\ve{X};[\ve{Z}_R,\ve{z}_D]) \right\Vert_1 \bigg] - w_{\rm KL}D_{\rm KL}\left(q(\ve{z}_R) \parallel p(\ve{z}_R)\right),
\end{equation}
where $p(\ve{z}_R) = \mathcal{N}(\ve{0},\beta^2\ve{I})$ is the Gaussian prior density for the random parameters $\ve{Z}_R$. The first term is the expected loss under $q(\ve{z}_R)$, using the L1-norm (equivalent to a multivariate Laplace likelihood) to maintain consistency with the MAE loss \eqref{eq: MAE} used in LSTM+MLP. The second term is the weighted negative KL divergence between the approximate density $q(\ve{z}_R)$ and the prior $p(\ve{z}_R)$. Since both densities are Gaussian, they form a conjugate exponential family pair, enabling the KL divergence to be evaluated efficiently in closed form. The weight $w_{\rm KL} = 10^{-3}$ balances the scale difference between these two terms in \eqref{eq: LSTM+BNN ELBO}. Intuitively, the first term ensures that the model fits the data effectively, while the second term maintains consistency with our prior beliefs. 

During training, the variational parameters $\ve{\mu}_R$, $\ve{\sigma}_R$, and deterministic parameters $\ve{z}_D$ are optimized simultaneously. For a detailed description of the LSTM+BNN training procedure, see Algorithm \ref{algo: Training procedure} in \ref{app: additional}.

To generate probabilistic predictions from the LSTM+BNN model, the following Monte Carlo sampling strategy is applied. An input sequence $\ve{X}^{(t)}$ is passed through the model in \eqref{eq: LSTM+BNN} multiple times. At each forward pass, a realization $\ve{z}_R$ of the random parameters $\ve{Z}_R$ is sampled from the approximate posterior distribution $q(\ve{z}_R)$ parameterized by the optimized $\ve{\mu}_R$ and $\ve{\sigma}_R$. This yields an array of predictions, from which the predictive mean and the standard deviation---a measure of predictive uncertainty---can be calculated.

\section{Experiments}\label{Experiments}

\subsection{Dataset}\label{Dataset}
The building-specific datasets were collected by a private company specializing in MPC for centrally-heated residential buildings. For a total of $S=100$ buildings, the datasets contain hourly time series data between the years 2017 and 2022. The indoor temperature data represent the averages of sensor measurements from apartments within each building. The space heating supply water temperature data are measured at the building-specific heating substations. The outdoor temperature and the solar irradiation data are from the professional weather data provider of the company, fetched for the buildings based on their GPS coordinates. The availability of all these data is an essential prerequisite for the practical implementation of the indoor temperature models covered in this work.

Since the heating season in northern climates spans September to May, only data from these months are considered. During the summer months, the heating systems are shut off, and therefore, the indoor temperature cannot be affected by controlling the supply water temperature of the space heating circuit.

\subsection{Model implementations}\label{Model implementations}
Given the unique thermodynamics of each building and its heating system, models are trained separately for each of the $S$ buildings. Thus, we end up having $S$ reference models, $S$ LSTM+MLP models, and $S$ LSTM+BNN models. All model formulations can be extended to any building with sufficient data on inputs and outputs, allowing the building-specific thermal dynamics to be expressed through model parameters learned from the data. Although the DL models are flexible enough to train a single global model for knowledge transfer between buildings with similar thermal behavior \cite{Hakkinen_2022}---an approach beneficial for buildings with insufficient data---our aim here is to develop building-specific models for cases with comprehensive datasets. In industrial heating MPC applications, one can continuously evaluate different models and the available datasets for different buildings and prediction circumstances, and dynamically choose the model with the best potential for each building and prediction scenario.

Details regarding the inputs and outputs of the reference model are covered in Section \ref{Reference method}. The state and parameter estimation of the reference model for each building is performed using the data of the respective building from the last 500 heating season days prior to September 2021.

To train the DL models, all available site-specific data prior to September 2021 are split into \textit{training} and \textit{validation} sets at a 9-to-1 ratio.
The models are implemented using software packages \texttt{PyTorch} \cite{Paszke_2019} and \texttt{torchbnn} \cite{Lee_2022} as auto-regressive predictors that work on an hourly domain. That is, although the models are used to predict multiple time steps into the future, in the training phase, they are fitted based on their 1-hour forward predictions \cite{Hakkinen_2022}.
Both DL model variants are trained using the Adam optimizer with an initial learning rate of $10^{-4}$, decayed by a factor of $\frac{1}{2}$ over 3 scheduled reductions. The number of training epochs is 400 for the LSTM+MLP and 800 for the LSTM+BNN. Both models are trained using a single batch, and no specific random seed is specified.

\subsection{Results}
\label{Results}

In the MPC application for which the models are developed, control sequences are optimized hourly based on indoor temperature predictions over a prediction horizon of length $H=48$.
Therefore, to maintain the system state at a target value and minimize overheating with MPC, we are particularly interested in the models' short-term predictive performance. However, long-term predictions over the full prediction horizon of length $H$ are crucial for achieving optimal control planning, especially when considering advanced optimization objectives (discussed in Section \ref{Intro}).

The \textit{test} set comprises data from October 2021 to April 2022. Within this date range, $T = 100$ evenly distributed hourly time instances $\ve{t}\in \mathbb{R}^T$ are selected to comprehensively cover the heating season conditions when testing the models. For each of the $S$ buildings, starting from the selected time instances $\ve{t}$, we generate $S \cdot T = 10000$ indoor temperature prediction sequences into a prediction horizon of length $H = 48$ using all three models. Thus, the $T_{{in}}$ predictions of a single model type are denoted as a matrix $\ve{\hat{Y}} \in \mathbb{R}^{(S\cdot T) \times H}$.

The $\Delta T_{{in}}$ predictions of the DL models are transformed into $T_{{in}}$ predictions by cumulatively adding the predicted differences to the latest indoor temperature measurement the model had available given a starting point $t \in \ve{t}$. Since $T_{{in}}$ is incorporated in the DL models' input variables (see Table \ref{tab: Input}), each prediction beyond the first one into the prediction horizon is affected by the model's own previous predictions. In the reference model, the same applies inherently due to the LSSM formulation.

Finally, we note that the predictions of the probabilistic LSSM and LSTM+BNN models refer to their respective predictive means. The Monte Carlo sampling strategy for generating probabilistic predictions from the partially stochastic LSTM+BNN is detailed in Section \ref{LSTM+BNN}. In all numerical experiments, we used $N_S=10$ posterior predictive samples, with the exception of Section \ref{sec: UQ}, for which $N_S=100$.

\subsubsection{Prior selection for the partially stochastic deep learning model}
\label{prior selection}

\begin{figure}[ht!]
    \centering
    \includegraphics[width=0.9\linewidth]{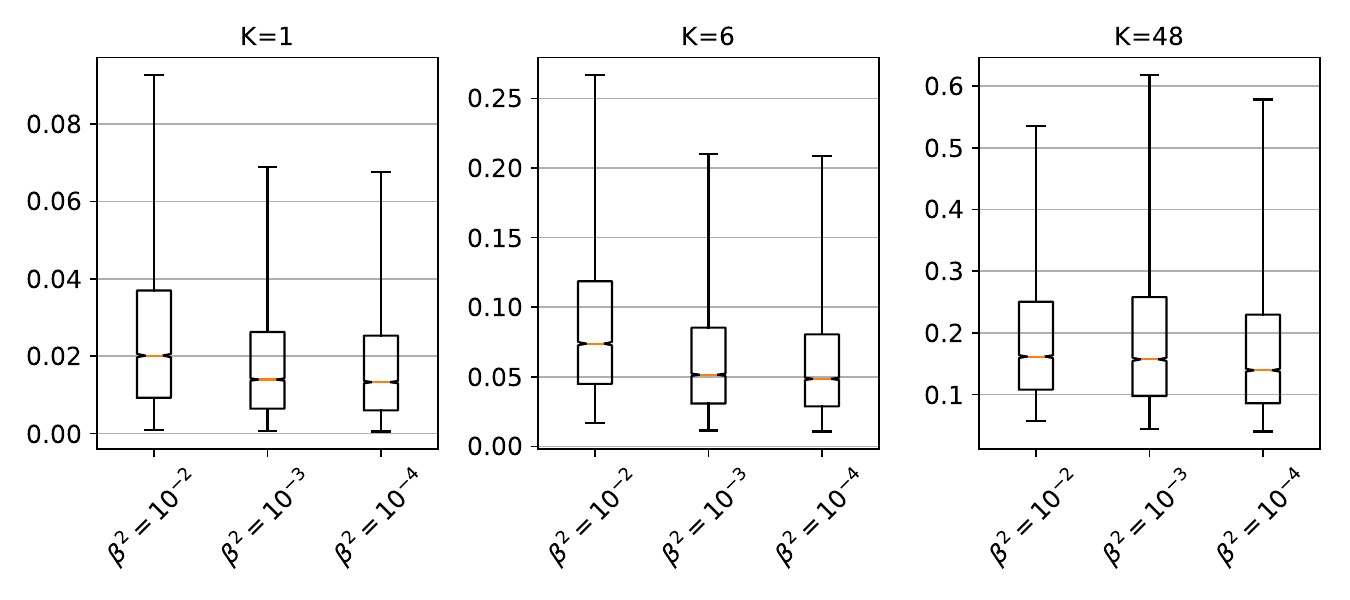}
    \caption{Performance of the LSTM+BNN with different prior variances $\beta^2$ as RMSE \eqref{RMSE} for different prediction horizon lengths $K\in[1, 6, 48]$. Orange lines are the RMSE distribution medians, while boxes and whiskers indicate the central $50\%$ and $95\%$ of the data, respectively. Note the different scales on the Y-axes.}
    \label{fig: Priors}
\end{figure}

The LSTM+BNN model is tested with three prior variances $\beta^2$: $10^{-2}$, $10^{-3}$, and $10^{-4}$.
Models trained with each prior variance are used to compute $\ve{\hat{Y}} \in \mathbb{R}^{(S\cdot T) \times H}$. We then evaluate the models' predictive performance into different horizon lengths $K\in[1, 6, 48]$ by calculating the Root Mean Squared Errors (RMSEs) w.r.t. $T_{{in}}$ measurements $\ve{Y} \in \mathbb{R}^{(S\cdot T) \times H}$ along the horizon lengths $K$ as
\begin{equation}
\label{RMSE}
    \ve{\epsilon}_{i}^{{hor}} = \sqrt{\frac{\sum_{j=1}^K \left( \ve{Y}_{i, j} - \ve{\hat{Y}}_{i, j} \right)^2}{K}}, \quad i = 1, \ldots, S \cdot T.
\end{equation}
The respective RMSE distributions are shown in Figure \ref{fig: Priors}. Notice the different scales on the Y-axes to facilitate easier visual comparison of the LSTM+BNN with different prior variances for each prediction horizon length $K$.

In addition to predictive performance, the value at which the KL divergence term in \eqref{eq: LSTM+BNN ELBO} converges during training must be considered.
Since this term measures the dissimilarity between the prior and the variational approximation, the closer we get to zero, the better. Although the LSTM+BNN with a prior variance of $10^{-4}$ has the best accuracy, the KL divergence term is 100 times larger compared to the models with prior variances of $10^{-2}$ and $10^{-3}$.
A small prior variance combined with a large KL divergence suggests that the prior is too restrictive.
This can lead to instability or overfitting, as the model may not effectively regularize its parameters.
Therefore, since the prior $\mathcal{N}(0,10^{-3}\ve{I})$ achieves both good predictive performance and a low value for the KL divergence term, it is selected for the LSTM+BNN employed in the comparative experiments of the study.

This empirical analysis provides a principled, data-driven justification for the selected prior variance of $10^{-3}$, balancing predictive accuracy with stable posterior regularization for the building types in our dataset. We note that the optimal value is inherently dependent on the data characteristics of the specific buildings under study. A comprehensive sensitivity analysis across a wider variety of building types and a deeper physical interpretation of these hyperparameters would be greatly facilitated by a hierarchical Bayesian framework, where the priors themselves are learned from the data. This is a promising direction for enhancing the model's generalizability and robustness, which we identify as an important objective for future work. 

Finally, Table \ref{tab: hyperparameters} summarizes all hyperparameters of the DL models employed in the further experiments of this study. 

\begin{table}[!ht]
    \centering
    \caption{Hyperparameters of the deep learning models.}
    \begin{tabular}{l| c c}
    \hline
        & LSTM+MLP & LSTM+BNN \\
        \hline
        Epochs & 400 & 800 \\
        Initial learning rate & \multicolumn{2}{c}{$10^{-4}$} \\
        Optimizer & \multicolumn{2}{c}{Adam}\\
        Hidden units & \multicolumn{2}{c}{1024 / 1024 / 512} \\
        KL weight, $w_{\rm KL}$ & - & $10^{-3}$ \\
        Prior variance, $\beta^2$ & - & $10^{-3}$ \\
        \hline
    \end{tabular}
    \label{tab: hyperparameters}
\end{table}

\subsubsection{Predictive performance} 

\begin{figure}[ht!]
     \centering
     \includegraphics[width=\linewidth]{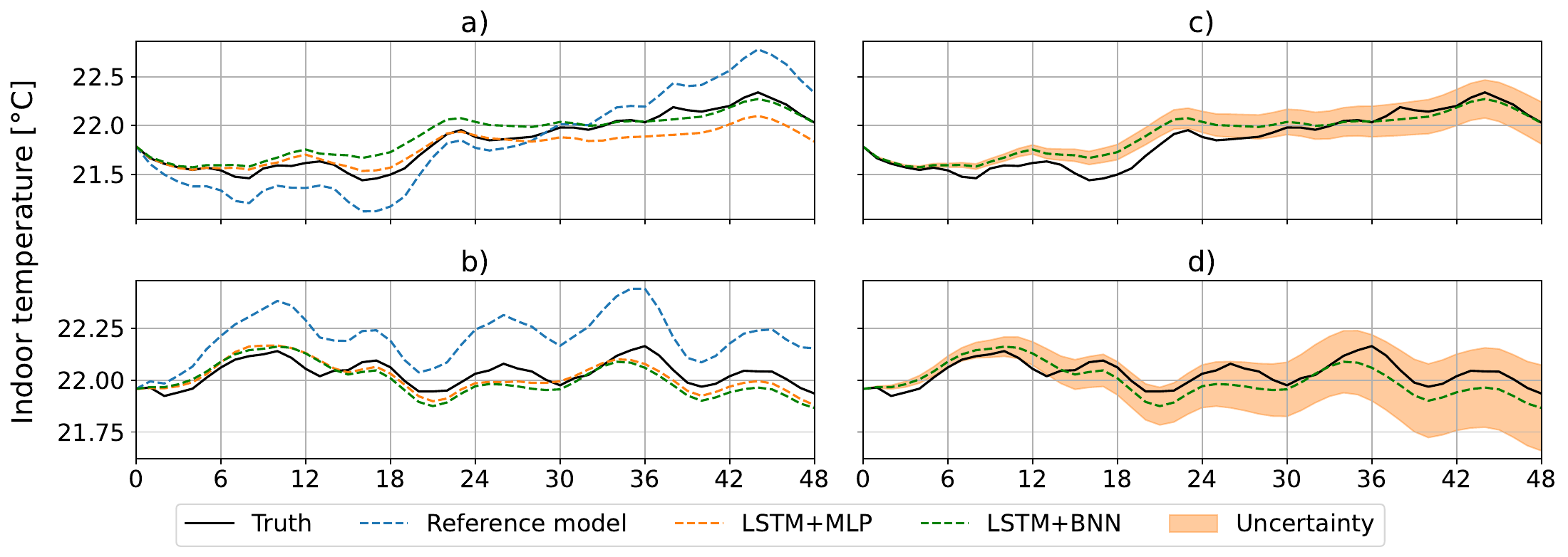}
     \caption{a) and b): 48-hour indoor temperature predictions for two different test samples. c) and d): Predictive uncertainties of the LSTM+BNN corresponding to the test samples shown in a) and b), respectively. Uncertainty is quantified as the cumulative standard deviation across each prediction's $N_S=10$ posterior predictive samples.}
     \label{fig: Predictions}
\end{figure}

For two test samples, Figure \ref{fig: Predictions} shows the predictions of all three models and the predictive uncertainties of the LSTM+BNN model over the 48-hour prediction horizon. Since the uncertainties of the successive LSTM+BNN model predictions are independent, the predictive uncertainty at each prediction horizon length $K\in[1, \ldots, 48]$ is calculated by accumulating the standard deviations of all predictions up to that point.

\begin{figure}[ht!]
    \centering
    \includegraphics[width=0.9\linewidth]{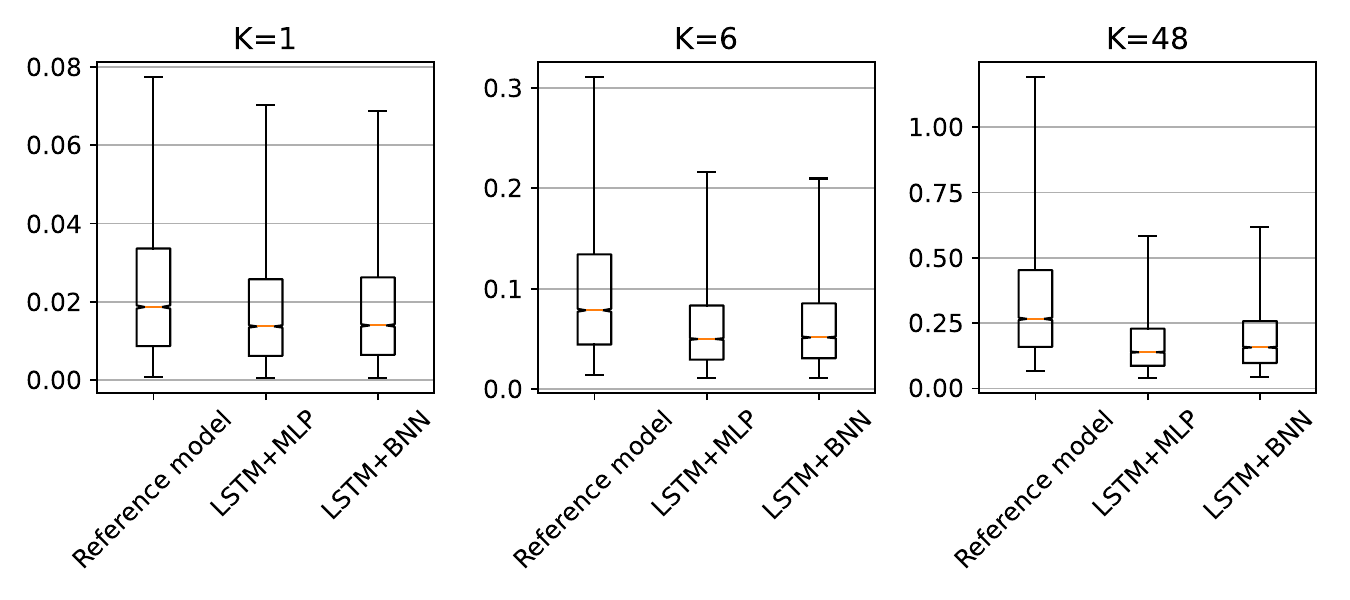}
    \caption{Performance of the three models as RMSE \eqref{RMSE} distributions into different prediction horizon lengths $K\in[1, 6, 48]$. Orange lines are the RMSE distribution medians, while boxes and whiskers indicate the central $50\%$ and $95\%$ of the data, respectively. Note the different scales on the Y-axes.}
    \label{fig: RMSE}
\end{figure}

To compare the predictive performance of the models across different buildings and heating season conditions, we use the same RMSE metric \eqref{RMSE} as we did when selecting the prior variance for the LSTM+BNN in Section \ref{prior selection}. The results are shown in Figure \ref{fig: RMSE}. Again, notice the different scales on the Y-axes to facilitate easier visual comparison of the models for each prediction horizon length $K\in[1, 6, 48]$. Both DL models outperform the reference for all three prediction horizon lengths, particularly for longer prediction horizons. The short-term predictions of the DL models are also significantly more accurate with a tighter RMSE distribution. The difference in predictive performance between the LSTM+MLP and LSTM+BNN models is minimal.

\begin{figure}[ht!]
    \centering
    \includegraphics[width=0.75\linewidth]{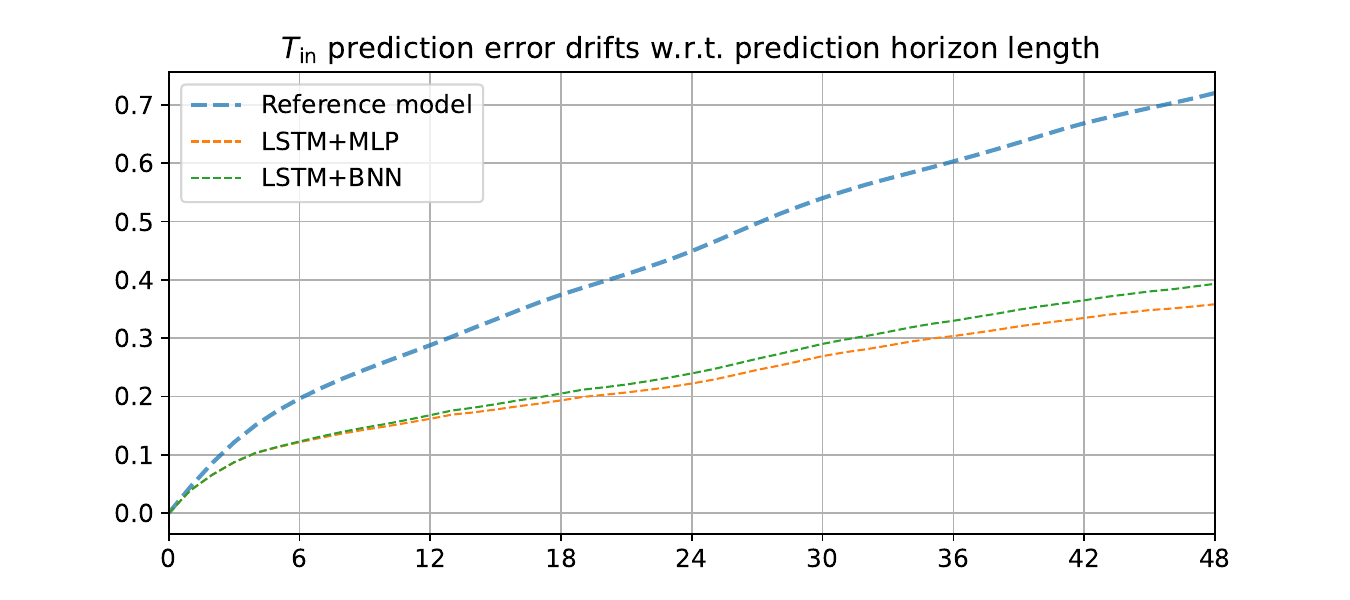}
    \caption{Average prediction drifts across the test samples \eqref{drift} with the three models.}
    \label{fig: Drift models}
\end{figure}

To evaluate how much the predictions $\ve{\hat{Y}} \in \mathbb{R}^{(S\cdot T) \times H}$ of each model drift from the measured truth $\ve{Y} \in \mathbb{R}^{(S\cdot T) \times H}$ when predicting into the prediction horizon of length $H=48$, we calculate the average drift over our test samples w.r.t. prediction horizon length using a similar metric to \eqref{RMSE}, but this time averaging along the test samples:
\begin{equation}
\label{drift}
    \ve{\epsilon}_{j}^{{dri}} = \sqrt{\frac{\sum_{i=1}^{(S \cdot T)} \left( \ve{Y}_{i, j} - \ve{\hat{Y}}_{i, j} \right)^2}{S \cdot T}}, \quad j = 1, \ldots, H.
\end{equation}
The average prediction drifts for all three models are visualized in Figure \ref{fig: Drift models}. The results show that the DL models achieve progressively better performance relative to the reference model as the prediction horizon lengthens. Over the test samples, the average performance of the DL models stays similar until the 6-hour horizon length, after which the deterministic LSTM+MLP performs slightly better.

\begin{figure}[ht!]
    \centering
    \includegraphics[width=0.6\linewidth]{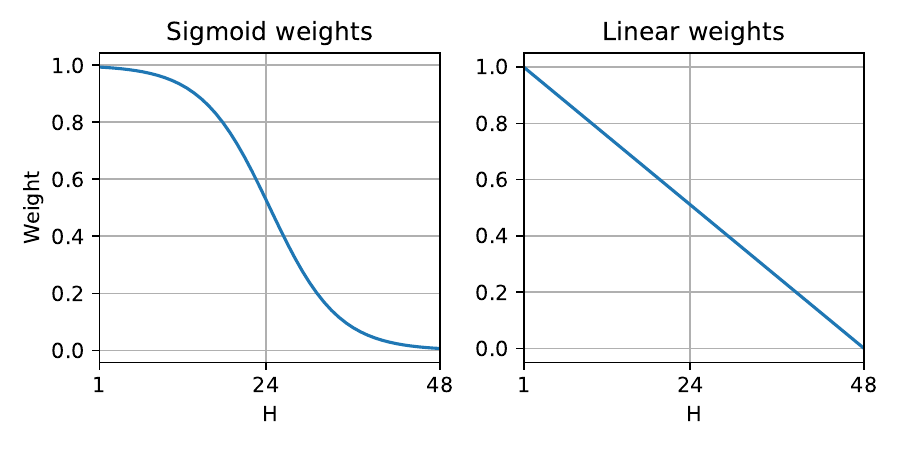}
    \caption{Sigmoid and linear functions used to calculate weighted predictive performance scores in Table \ref{tab: Weighted RMSE}.}
    \label{fig: Weights}
\end{figure}

As discussed earlier, our interest lies more in short-term predictive accuracy than long-term. Hence, to compare the performance of the models using a single metric aligned with what we are interested in, the prediction drift series visualized in Figure \ref{fig: Drift models} is averaged using linear- and sigmoid-shaped weight functions. These functions give decreasing weights for the RMSE values w.r.t. prediction horizon length. The sigmoid function gives more weight to the accuracy of the short-term prediction compared to the linear function (Figure \ref{fig: Weights}). The performance scores calculated this way are shown in Table \ref{tab: Weighted RMSE} for all three models.

\begin{table}[ht!]
    \centering
    \caption{Predictive performance scores for the three models calculated by averaging the prediction drifts w.r.t. prediction horizon length (cf. Figure \ref{fig: Drift models}) using different weight functions.}
    \begin{tabular}{c c c c}
    \hline
       Model & Unweighted & Sigmoid & Linear \\
       \hline
       Reference  & 0.436 & 0.145 & 0.161\\
       
       LSTM+MLP & 0.225 & 0.079 & 0.086\\
       
       LSTM+BNN & 0.247 & 0.084 & 0.093 \\
       \hline
    \end{tabular}
    \label{tab: Weighted RMSE}
\end{table}

\subsubsection{Uncertainty quantification}
\label{sec: UQ}

\begin{figure}[ht!]
    \centering
    \includegraphics[width=0.5\linewidth]{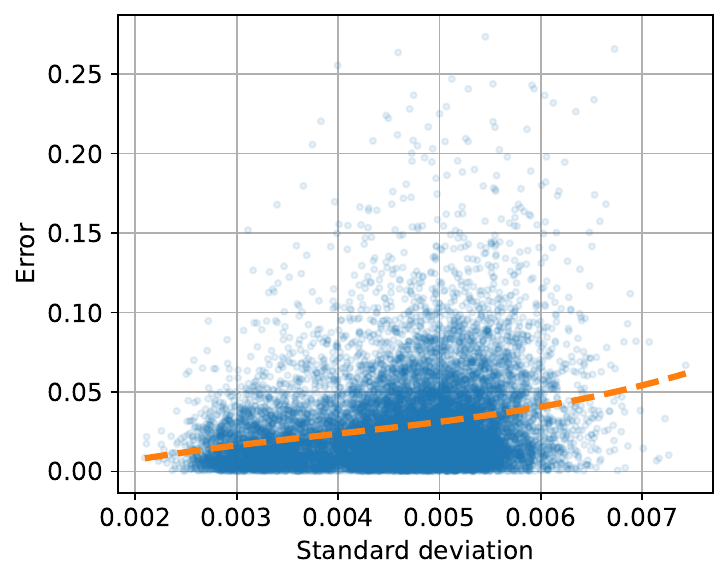}
    \caption{Predictive uncertainties of the LSTM+BNN models (as standard deviations) and errors of the respective predictive means (as MAEs) for the $S \cdot T$ test samples into a 1-hour prediction horizon. The predictive mean and standard deviation were calculated from $N_S=100$ posterior predictive samples for each test sample.}
    \label{fig: Std_Error}
\end{figure}

Figure \ref{fig: Std_Error} illustrates the average increase in prediction error of the LSTM+BNN models with respect to the predictive uncertainty. Although the LSTM+BNN performs slightly worse in terms of pure predictive performance compared to its deterministic variant, LSTM+MLP, (see Figure \ref{fig: Drift models} and Table \ref{tab: Weighted RMSE}), it offers a key advantage in having the ability to quantify the uncertainty present in its predictions. This transparency regarding model reliability is highly beneficial when it comes to dynamically selecting the best model for each building and prediction scenario in industrial heating MPC applications.

\subsubsection{Computational costs}

The primary consideration for deploying DL models, compared to the reference model, is their computational cost (see Table \ref{tab: times}). In the industrial heating MPC application for which the models are developed, the workflows for model training, model selection, and control optimization---where models are used for predictions---are separate and parallelizable computational tasks. This architecture is key to scalability. Currently, for each building, the reference model training is scheduled for every other hour, the model selection for once a week, and the control optimization is performed hourly.

\begin{table}[ht!]
    \centering
    \caption{Computational costs of the models. The training time is the average time taken to train a model for a single building from scratch. The prediction time is the average time it took to generate a single 48-hour prediction sequence.}
    \begin{tabular}{c c c}
    \hline
         Model &  Training time (s) & Prediction time (s)\\
         \hline
        Reference & 5 & 0.008\\
        LSTM+MLP & 180 & 0.17\\
        LSTM+BNN  & 360 & 0.31\\
        \hline
    \end{tabular}
    \label{tab: times}
\end{table}

For buildings with limited data, such frequent model updates are typically necessary, making the reference model with computationally efficient posterior inference the preferred choice. However, for buildings with comprehensive datasets where the DL models are applicable, the training frequency can be significantly reduced---for instance, to once per week. Furthermore, the reported training times (Table \ref{tab: times}) are for training the models from scratch. In practice, retraining can be initialized from the previous model's parameters, drastically reducing the number of iterations and computational cost required for convergence. The proposed LSTM+BNN model is intended for MPC of these data-rich buildings, where its superior accuracy justifies the computational expense. Therefore, while the aggregate training cost across a large portfolio is a defined limitation, the appropriate retraining scheduling and initialization enabled by the scalable architecture prevent it from being a prohibitive barrier to industrial deployment.

In the control optimization task, the space heating supply temperature control sequence for the next $H$ hours is initialized based on the most recent optimization results. This sequence is then optimized according to various objectives, taking into account recent observations of the system state, current weather forecasts, and model predictions. Due to the effective initialization, the optimization process typically concludes after just a few iterations. Consequently, the slightly longer prediction time of the LSTM+BNN compared to the reference model does not hinder its practical use in the control optimization loop.

\section{Conclusions and future work}
\label{Conclusions}

We developed a partially stochastic DL architecture, termed LSTM+BNN, for building-specific indoor temperature modeling that incorporates built-in UQ. The method was compared against its deterministic DL variant, LSTM+MLP, and an industry-proven physics-based Bayesian LSSM. Rigorous comparative experiments were conducted using a large dataset collected from 100 centrally-heated residential buildings, extensively covering different heating season conditions.

Both of the DL models outperformed the reference model significantly in short- and long-term predictions. The improved performance of the DL models comes from their ability to capture complex nonlinear thermodynamics, including thermal lag effects (heat storage in building structures) and solar energy gain variations. 
While the DL variants showed similar accuracy, the partially stochastic version offers transparency by providing a way of quantifying the uncertainty present in its predictions. This feature is highly beneficial for integrating the data-driven model into real-world heating MPC solutions, as model suitability for a building and prediction scenario can be dynamically evaluated prior to each control optimization.

Although our validation was conducted on Nordic residential buildings, the adoption of a first-order nodal view for the modeled system, combined with the data-driven, building-specific nature of the covered methods, suggests strong potential for generalizability across building types (e.g., commercial, offices, mixed) and climate zones. Both the reference and proposed model learn parameters directly from building-specific data, allowing them to naturally adapt to different contexts.

The higher computational cost of the LSTM+BNN model is a defined limitation, but we have justified it as manageable through appropriate retraining scheduling and initialization, thus not impeding its practical MPC integration for data-rich buildings.
Therefore, integrating the partially stochastic LSTM+BNN into heating MPC solutions---especially for buildings with comprehensive datasets---has great potential to improve control accuracy, thereby enhancing heating control in terms of energy efficiency, thermal comfort, and success with more advanced optimization objectives.

The deployment of an industrial-grade MPC solution in a real-life application is far from trivial and requires significant infrastructure beyond the core model, including data connectivity, storage, robust implementation of control optimization, monitoring, handling connection issues, etc. This work provides a critical building block for such systems by introducing and validating a high-performance, uncertainty-aware model. The integration of this component into a full, robust industrial system remains a separate, important engineering challenge.

Finally, the results of this work motivate and lay the groundwork for further studies on improving probabilistic indoor temperature modeling in the context of heating MPC. Hence, we leave as future work the following:
\begin{itemize}
    \item[(i)] Understanding carefully which of the learned physical relationships from the data contribute to the enhanced modeling ability of data-driven models, i.e., identifying what is missing from the existing reference model formulation and how it could be improved. For instance, methods like \cite{Brunton_2016, Cranmer_2023} could be used to find interpretable, functional formulations for the input-output mappings learned by the data-driven models.
    \item[(ii)] Increasing the inductive bias of the data-driven models by incorporating domain knowledge into the training process, i.e., finding the balance between theory- and data-driven modeling \cite{Raissi_2019, Bastek_2025, Chen_2019, Rackauckas_2021}. This could enhance the generalization ability, especially with limited or noisy training data.
    \item[(iii)] Heating dynamics of buildings exhibit seasonal patterns, slow trends, and sudden shifts (e.g., from physical modifications). Future work could also focus on adaptive methods that balance learning from new data and discarding outdated information, possibly via time-evolving parameters \cite{Rimella2025}.
    \item[(iv)] Enhancing the robustness and generalizability of the proposed DL framework. This includes developing more automated and principled methods for hyperparameter selection, such as hierarchical Bayesian modeling, to adapt the prior variances and KL weighting to different building types and data regimes without manual tuning. Furthermore, architectural optimizations for computational efficiency will be crucial for large-scale deployment.
    \item[(v)] The logical and necessary next step is the practical implementation and testing within a closed-loop MPC framework. Future work should focus on case studies to evaluate the interaction of the proposed model with different optimization objectives, ensuring control stability. Furthermore, to guarantee robust control actions, the physics-consistency of the model predictions (e.g., monotonicity with respect to certain inputs) must be rigorously studied and enforced. These steps are critical for the large-scale industrial deployment of advanced data-driven models like the one proposed.
\end{itemize}

\section{Acknowledgments}
We would like to give special thanks to Jaakko Luttinen for his work on the reference method and to Alexander Ilin for his involvement in the conceptualization of the project. Furthermore, we thank Danfoss Leanheat for providing the data used in the study.

This work was supported by the Finnish Ministry of Education and Culture’s Pilot for Doctoral Programmes (Pilot project Mathematics of Sensing, Imaging and Modelling) and Research Council of Finland (Flagship of Advanced Mathematics for Sensing and Imaging and Modeling grant 359183, Centre of Excellence of Inverse Modelling and Imaging grant 353095). The research of Jana de Wiljes has been partially funded by the Deutsche Forschungsgemeinschaft (DFG)- Project-ID 318763901 - SFB1294. Furthermore, this project has received funding from the European Union under the Horizon Europe Research \& Innovation Programme (Grant Agreement no. No 101188131 UrbanAIR). Views and opinions expressed are however those of author(s) only and do not necessarily reflect those of the European Union. Neither the European Union nor the granting authority can be held responsible for them.

\addcontentsline{toc}{section}{References}
\bibliographystyle{model1-num-names}
\bibliography{bibliography.bib}

\newpage
\appendix

\section{}
\label{app: additional}

\begin{algorithm}[!ht]
\caption{Detailed LSTM+BNN training procedure.}
\label{algo: Training procedure}
\begin{algorithmic}
\State $\ve{X} \in \mathbb{R}^{N \times L \times M}, \ \Delta\ve{y}\in\mathbb{R^N}$ \Comment{Dataset}
\State $\beta^2 \gets 10^{-3}$, $w_{\rm KL} \gets 10^{-3}$ \Comment{Set prior variance and KL weight}
\State $p(\ve{z}_R) \gets \mathcal{N}(\ve{0},\beta^2\ve{I})$ \Comment{Define BNN prior}
\State $\ve{\mu}_R \gets \ve{0}$, $\ve{\sigma}_R \gets \ve{10^{-3}}$, $\ve{z}_D \gets \ve{0}$ \Comment{Initialize model parameters}
\State $q \gets \mathcal{N}(\ve{\mu}_R, \ve{\sigma}_R^2\ve{I})$ \Comment{Set variational density}
\State $\mathcal{L}_{\text{val,min}} \gets \infty$ \Comment{Initialize minimum validation loss}

\Procedure{Training}{$\ve X$}

    \State $\ve{X} = \{\ve X_{\text{train}}\in \mathbb{R}^{N_{\text{train}} \times L \times M},\ve X_{\text{val}}\in \mathbb{R}^{N_{\text{val}} \times L \times M}\}$
     \State $\Delta \ve{y} = \{\Delta \ve{y}_{\text{train}}\in \mathbb{R}^{N_{\text{train}}}, \Delta \ve{y}_{\text{val}}\in \mathbb{R}^{N_{\text{val}}}\}$
    \For{$i = 1$ to \texttt{epochs}}
        \State Draw $\ve{z}_R\sim q$ for each $\ve X_{\text{train}}^{(t)}$ 
        \State $\Delta  \hat{\ve y}_{\text{train}} \gets \Psi(\ve X_{\text{train}}; [\ve{z}_R,\ve{z}_D])$ \Comment{See equation \eqref{eq: LSTM+BNN}}
        \State With $\Delta \hat{\ve y}_{\text{train}}, \Delta \ve y_{\text{train}}, q$ evaluate ELBO as \emph{training loss}, $\mathcal{L}_{\text{train}}$ \Comment{See equation \eqref{eq: LSTM+BNN ELBO}}
        \State Back-propagate $\mathcal{L}_{\text{train}}$ to compute gradients w.r.t. $\ve{\mu}_R, \ve{\sigma}_R, \ve{z}_D$
        \State Update $\ve{\mu}_R, \ve{\sigma}_R, \ve{z}_D$ based on learning rate and the gradients \Comment{Adam optimizer}
        \State $q \gets \mathcal{N}(\ve{\mu}_R, \ve{\sigma}^2_R\ve{I})$ \Comment{Update variational density}
        \State Draw $\ve{z}_R\sim q$ for each $\ve X_{\text{val}}^{(t)}$  
        \State $\Delta  \hat{\ve y}_{\text{val}} \gets \Psi(\ve X_{\text{val}}; [\ve{z}_R,\ve{z}_D])$ \Comment{See equation \eqref{eq: LSTM+BNN}}
        \State With $\Delta \hat{\ve y}_{\text{val}}, \Delta \ve y_{\text{val}}, q$ evaluate ELBO as \emph{validation loss}, $\mathcal{L}_{\text{val}}$ \Comment{See equation \eqref{eq: LSTM+BNN ELBO}}
        \If{ $\mathcal{L}_{\text{val}}<\mathcal{L}_{\text{val,min}}$}
        \State Save model with current parameters $\ve{\mu}_R, \ve{\sigma}_R, \ve{z}_D$ as \emph{the best model}, $\Psi_\text{best}$
        \State  $\mathcal{L}_{\text{val,min}} \gets \mathcal{L}_{\text{val}}$
    \EndIf
    \EndFor

    \State \textbf{return} $\Psi_\text{best}$
    
\EndProcedure
\end{algorithmic}
\end{algorithm}

\end{document}